\documentclass{aa}
\usepackage[modulo,switch]{lineno}
\usepackage{graphicx}
\usepackage{natbib}
\usepackage{txfonts}
\begin{document}
\newcommand{\meandnu} {\langle\Delta\nu\rangle}
   \title{Asteroseismic surface gravity for evolved stars}

   \author{S. Hekker\inst{1,2} \and Y. Elsworth\inst{2}  \and B. Mosser\inst{3} \and T. Kallinger\inst{4} \and Sarbani Basu\inst{5} \and W.J. Chaplin\inst{2}  \and D. Stello\inst{6}}

\offprints{S. Hekker, \\
                    email: S.Hekker@uva.nl}
   \institute{Astronomical institute `Anton Pannekoek', University of Amsterdam, Science Park 904, 1098 XH, Amsterdam, the Netherlands
   \and University of Birmingham, School of Physics and Astronomy, Edgbaston, Birmingham B15 2TT, United Kingdom
   \and LESIA, UMR8109, Universit\'e Pierre et Marie Curie, Universit\'e Denis Diderot, Observatoire de Paris, 92195 Meudon Cedex, France
    \and  Institute for Astronomy, University of Vienna, T\"urkenschanzstrasse 17, A-1180 Vienna, Austria
   \and  Department of Astronomy, Yale University, P.O. Box 208101, New Haven CT 06520-8101, USA 
   \and Sydney Institute for Astronomy (SIfA), School of Physics, University of Sydney, NSW 2006, Australia\\
         }

   \date{Received ; accepted}


  \abstract
   {Asteroseismic surface gravity values can be of importance in determining spectroscopic stellar parameters. The independent $\log(g)$ value from asteroseismology can be used as a fixed value in the spectroscopic analysis to reduce uncertainties due to the fact that $\log(g)$ and effective temperature can not be determined independently from spectra. Since 2012, a combined analysis of seismically and spectroscopically derived stellar properties is  ongoing for a large survey with SDSS/APOGEE and \textit{Kepler}. Therefore, knowledge of any potential biases and uncertainties in asteroseismic $\log(g)$ values is now becoming important.}
   {The seismic parameter needed to derive $\log(g)$ is the frequency of maximum oscillation power ($\nu_{\rm max}$). Here, we investigate the influence  of $\nu_{\rm max}$ derived with different methods on the derived $\log(g)$ values. The large frequency separation between modes of the same degree and consecutive radial orders ($\Delta\nu$) is often used as an additional constraint for the determination of $\log(g)$. Additionally, we checked the influence of small corrections applied to $\Delta\nu$ on the derived values of $\log(g)$. }
   {We use methods extensively described in the literature to determine $\nu_{\rm max}$ and $\Delta\nu$  together with seismic scaling relations and grid-based modeling to derive $\log(g)$. }
   {We find that different approaches to derive oscillation parameters give results for $\log(g)$ with small, but different, biases for red-clump and red-giant-branch stars. These biases are well within the quoted uncertainties of $\sim0.01$ dex (cgs). Corrections suggested in the literature to the $\Delta\nu$ scaling relation have no significant effect on $\log(g)$. However somewhat unexpectedly, method specific solar reference values induce biases of the order of the uncertainties, which is not the case when canonical solar reference values are used.}
   {}

   \keywords{asteroseismology -- stars: fundamental parameters -- stars: oscillations}
   \maketitle
%

\section{Introduction}
With the current wealth of data, the community has great opportunities to improve the knowledge of stellar parameters. One of the important characteristics of stars is their surface gravity ($g$). Surface gravity can be determined in several independent ways, such as from stellar spectra or asteroseismology, i.e., from the intrinsic oscillations of stars.

Several studies have explored the accuracy of asteroseismic $\log(g)$ values. For main-sequence and subgiant stars the accuracy of the determined asteroseismic $\log(g)$ has been investigated by comparisons with $\log(g)$ values from classical spectroscopic methods \citep[e.g.][]{morel2012} and independent determinations of radius and mass \citep[e.g.][]{creevey2012,creevey2013}. These studies found good agreement between the gravities inferred from asteroseismology and spectroscopy, which supports the use of asteroseismic $\log(g)$. For more evolved stars -- the subject of this paper -- a small sample has been investigated by \citet{morel2012} which for $\log(g)$ values down to 2.5 dex (cgs) also showed good agreement. \citet{Thygesen2012} showed a comparison of spectroscopic and asteroseismic $\log(g)$ values for 81 low-metallicity stars with $\log(g)$ down to 1.0 dex (cgs). Also for this sample there is good agreement between the values supporting the use of asteroseismic $\log(g)$ determinations for evolved stars.

The principle of deriving surface gravity from stellar spectra is well understood in general. In practice, however,  the results depend on the specific technique used, i.e., ionization balance, line fitting or isochrone fitting, and their exact implementation. These differences can easily result in differences of the order of 0.2 dex \citep[e.g.,][]{hekker2007,morel2012} in $\log(g)$. A significant contribution to this uncertainty is due to the correlation between $\log(g)$ and effective temperature in the spectral analysis. One way to reduce the uncertainties caused by this correlation is to fix one of the parameters to an independently determined value. Asteroseismology provides such a route to determine $\log(g)$ in an independent way.

The quoted uncertainties of the asteroseismic $\log(g)$ are often an order of magnitude lower than those quoted in spectroscopic analyses, indicating more precise values. Indeed the high positive correlation between mass and radius leads to very small uncertainties in $M/R^2$ and hence in $\log(g)$. \citet{gai2011} showed that an asteroseismic $\log(g)$ can be obtained precisely and accurately with both direct and grid-based methods and that the result is largely model independent.

Over the past few years the number of stars with detected solar-like oscillations has increased considerably, from a few to over ten-thousand. For these large number of stars it is possible to derive an asteroseismic $\log(g)$ using global oscillation parameters, $\nu_{\rm max}$ (frequency of maximum oscillation power) and $\Delta \nu$ (large frequency spacing between modes of the same degree and consecutive orders). The potential of this was recognized in the field \citep[e.g.][]{gai2011} and several studies concerning the precision and accuracy of the asteroseismic $\log(g)$ values have been carried out. Two methods are generally used: 
\begin{itemize}
\item direct method: $\log(g)$ is computed from $\nu_{\rm max}$ from the scaling with the acoustic cut-off frequency $\nu_{\rm max} \propto g/\sqrt{T_{\rm eff}}$ \citep{brown1991,kjeldsen1995}. 
\item grid-based modeling: characteristics ($\log(g)$ in our case) of stars are determined by searching among a grid of models to get a "best model" for a given set of observables ($\Delta\nu$, $\nu_{\rm max}$, effective temperature ($T_{\rm eff}$) and preferably metallicity ([Fe/H])).
\end{itemize}
In the direct method it is implicitly assumed that all values for $T_{\rm eff}$ are possible for a star of a given mass and radius. However, the equations of stellar structure and evolution tell us that for a given mass and radius only a narrow range of temperatures are allowed. This is taken into account explicitly in the grid-based modeling since the grid is constructed by solving the equations of stellar structure and evolution.

We note here that it is also possible to compute $\log(g)$ and other parameters for a star from the individual frequencies, which are then directly compared with model predictions. In this case we need not rely on scaling relations. This route is however much more computationally intensive because frequencies need to be calculated for a dense grid of models, near surface effects play a more prominent role and we have to deal with the added complication of rotation and mixed gravity-pressure modes. Therefore, to determine $\log(g)$ for large samples of stars the use of seismic scaling relations for $\Delta\nu$ and $\nu_{\rm max}$ are currently preferred.
These seismic scaling relations relate the stellar mass, radius and effective temperatures with the observed global oscillation parameters: the frequency of maximum oscillation power ($\nu_{\rm max}$) and the frequency separation between modes of the same degree and consecutive orders ($\Delta\nu$). These scalings are performed with respect to solar reference values. The actual values of these solar references are debated \citep[e.g.][]{mosser2013} and discussed further in Sect. 2.3 and 3.4.

The validity of the scaling relations for $\nu_{\rm max}$ and $\Delta\nu$ for stars from the zero-age main sequence to the tip of the red giant branch are tested by e.g.~\citet{stello2008,stello2009,white2011} in a comparison with models. \citet{white2011} find that the scaling relation for $\Delta\nu$ is valid within $\sim$2\% with a dependence on effective temperature. The accuracy of the observed $\nu_{\rm max}$ and $\Delta\nu$ is such that this bias in the scaling relations is significant and has to be taken into account. This can either be done by using the equation suggested by \citet{white2011}, or by recalibration of the scaling relations as is done for cluster stars \citep[also using inferences from models,][]{miglio2012}. The inaccuracy in the scaling relations can also be accounted for in the uncertainties in $\log(g)$.

The computation of an asteroseismic $\log(g)$ requires the observations of global seismic parameters $\nu_{\rm max}$ and preferably also $\Delta \nu$. There are existing methods implemented in a range of algorithms to determine these global seismic parameters.  In these methods $\Delta\nu$ is computed as the mean large frequency separation over different frequency ranges. Throughout the paper we will refer to this quantity as the large separation. The consistency of and differences between the global seismic parameters have been studied by \citet{hekker2011,hekker2012} for red giant stars. These comparison studies show that the results for $\Delta\nu$ from different methods can be significantly different, depending on the evolutionary status of the star. This effect was not evident for $\nu_{\rm max}$, possibly due to the larger fractional uncertainties on this parameter. 

In this study we investigate the influence of the use of different global seismic parameters and methods on the determination of $\log(g)$. 
The study is driven by the large scale spectroscopic survey that is currently being conducted by the SDSS collaboration together with the \textit{Kepler} Asteroseismic Science Consortium with the APOGEE near-infrared spectrograph mounted on a 2.5 Ritchey-Chretien altitude-azimuth telescope located at Apache Point Observatory, New Mexico, USA. For this survey it is intended to determine spectroscopic effective temperatures and metallicities using asteroseismic surface gravities.

\begin{figure}
\begin{minipage}{\linewidth}
\centering
\includegraphics[width=\linewidth]{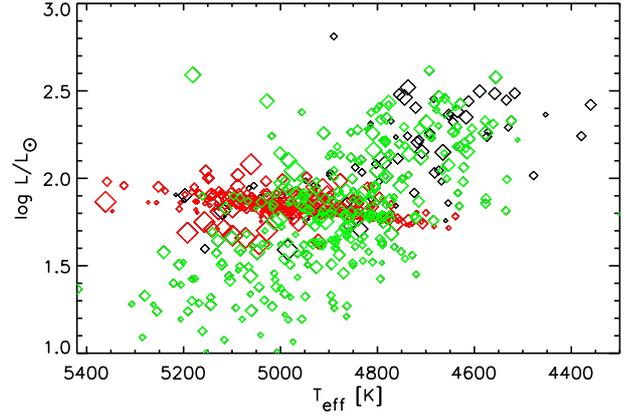}
\end{minipage}
\caption{Hertzsprung Russell diagram for stars used in this survey using SDSS temperatures calibrated with the infrared flux method \citep{pinsonneault2012} and luminosities computed using the asteroseismic radii computed using the OCT method and BaSTI models (see Sect. 3). Red-clump stars, red-giant branch stars and stars of unknown evolutionary phase are shown in red, green and black, respectively. The symbol sizes are proportional to the derived asteroseismic masses of the targets.}
\label{HRD}
\end{figure}

\section{Seismic scaling relations for $\nu_{\rm max}$ and $\Delta\nu$}

Seismic scaling relations for $\nu_{\rm max}$ and $\Delta\nu$ \citep{brown1991,kjeldsen1995} are used to relate the global oscillation properties of solar-like oscillators with stellar parameters mass ($M$), radius ($R$) - thus surface gravity ($g$) and mean density ($\bar{\rho}$) - and effective temperature ($T_{\rm eff}$) with all parameters expressed in solar values:

\begin{equation}
\nu_{\rm max} \approx \frac{M}{R^2\sqrt{T_{\rm eff}}} \approx \frac{g}{\sqrt{T_{\rm eff}}}
\label{numax}
\end{equation}
\begin{equation}
\Delta\nu \approx \sqrt{\frac{M}{R^3}} \approx \sqrt{\frac{g}{R}} \approx \sqrt{\bar{\rho}}
\label{dnu}
\end{equation}
These scaling relations are used with respect to solar values.

We discuss previous investigations of possible inaccuracies or biases in the scaling relations in more detail, followed by a discussion on the solar reference values.

\subsection{$\nu_{\rm max}$ scaling relation}
The scaling relation for $\nu_{\rm max}$ (Eq.~\ref{numax}) is an empirical relation in which $\nu_{\rm max}$ scales with the acoustic cut-off frequency \citep{brown1991}. The validity of this relation has recently been tested theoretically by \citet{belkacem2011}. They find a relation between the frequency at which the mode lifetime forms a plateau, i.e, $\nu_{\rm max}$, and the acoustic cut-off frequency, with a coefficient that depends on the ratio of the Mach number of the exciting turbulence to the third power to the mixing-length parameter.  So far the relation between this plateau and $\nu_{\rm max}$ has not been put on a theoretical basis. Nevertheless, this result is an important step towards understanding the underlying physics of the $\nu_{\rm max}$ scaling relation. However, at this stage, the practical difficulties to estimate the Mach number in the upper stellar envelopes implies that it is difficult to use these ideas to predict possible biases or inaccuracies in the $\nu_{\rm max}$ scaling relation.

\subsection{$\Delta\nu$ scaling relation}
From stellar models, \citet{white2011} showed that proportionality between the mean density of the star and the large frequency separation squared (Eq.\ref{dnu}) shows discrepancies of a few percent level for stars evolving from the Zero Age Main Sequence (ZAMS) up towards the tip of the red giant branch, with a clear correlation with the effective temperature. The amount of the discrepancy can be as large as 2-3\% in either a positive or negative sense dependent on the temperature of the star. 
It is important to understand whether this correction has a significant effect on the determined values of $\log(g)$.
We tested this and find that the difference between the results with and without these corrections are not significant (see Sect. 5).

\citet{miglio2012} have investigated the accuracy of the $\Delta\nu$ scaling relation for stars on the red-giant branch (RGB) and in the red clump (RC). They find that the sound speed in the RC model of 1.2 M$_{\odot}$ is on average higher (at a given fractional radius) than that of the RGB model of the same mass and radius. The main reason being the different temperature profile in the two models. \citet{miglio2012} note that while the largest contribution to the overall difference originates in the deep interior, near-surface regions ($r/R \gtrsim 0.9$) also contribute (by 0.8 per cent) to the total 3.5 per cent difference in total acoustic radius. This percentage is expected to be mass-dependent and to be larger for low-mass stars, which have significantly different internal structure when ascending the RGB compared to when they are in the core He-burning phase. \citet{miglio2012} did not derive an accurate theoretical correction of the $\Delta\nu$ scaling. The suggested change in the $\Delta\nu$ scaling relation is however of the same order as mentioned by \citet{white2011}. Because these changes caused a difference in $\log(g)$ well below the uncertainties (see Sect. 5), we do not expect significant impact on the determined $\log(g)$ values from the effect mentioned by \citet{miglio2012}  and therefore we do not investigate this further.

\citet{mosser2013} state that using the value of the large separation around $\nu_{\rm max}$ is a proxy only and that the solar reference value in Eq.~\ref{dnu} should be the asymptotic value. We comment further on this in Sect. 2.3.

\subsection{Solar reference values}
The scaling relations (Eq.~\ref{numax}~and~\ref{dnu}) are expressed in terms of the relevant solar reference values. Changing the solar reference value for $\nu_{\rm max}$ will induce an offset in $\log(g)$ proportional to the logarithm of their ratio. This means that for example changing the solar reference for $\nu_{\rm max}$ from 3050~$\mu$Hz \citep[e.g.][]{kjeldsen1995} to 3120~$\mu$Hz \citep[e.g.][]{kallinger2010} will induce a change in $\log(g)$ of about 0.02 dex (cgs) which is significant.

Additionally, \citet{mosser2013} argue that using the observed solar values in the scaling relations (Eq~\ref{numax} and \ref{dnu}) is not actually correct and that this would introduce biases  of a few percent in $\Delta\nu$. They suggest that one should be using the asymptotic value of $\Delta \nu$ valid at high order modes (higher order than the observed modes). Using this paradigm the reference values derived  become $\Delta \nu = 138.8~\mu$Hz and $\nu_{\rm max} = 3106~\mu$Hz \citep{mosser2013}. These are valid for stars with masses below 1.3~M$_{\odot}$ and effective temperatures between 6500 and 5000 Kelvin reflecting the range of stars for which the scaling relations are most reliable \citep{white2011}.

In this work we did not implement the asymptotic solar reference values. Firstly, as stated by the authors the change in $\Delta\nu$ of the observed star and the reference $\Delta\nu$ are of the same order and the net effect of these changes on the derived $\log(g)$ is small. Even if this imlies a few percent change in $\Delta\nu$, the tests with the \citet{white2011} corrections show that the effect on $\log(g)$ is negligible. Secondly, no asymptotic solar reference values are available in the temperature range of the red giants. 

\section{Determination of surface gravity}

\subsection{Data}
We  perform this study for the same sample of stars as used by \citet{hekker2012} for which there was agreement in the global oscillation parameters obtained by the different methods and for which there are results from three methods (see Sect. 3.2). This resulted in a list of 707 red giants. For a subset of these stars we know their evolutionary phases determined from period spacings of mixed modes \citep{beck2011,bedding2011,mosser2011mm} and the phase shift of the central radial mode \citep{kallinger2012}. Their locations in an H-R diagram are shown in Fig.~\ref{HRD}. For these stars we use \textit{Kepler} timeseries corrected for instrumental effects in the way described  by \citet{garcia2011}. 
For the effective temperatures we use the SDSS temperatures calibrated with the infrared flux method \citep{pinsonneault2012}. The evolutionary phases are determined from different methods, i.e., period spacings of mixed modes \citep{bedding2011,mosser2011mm, mosser2012mm, stello2013} and phase shift of the central radial mode \citep{kallinger2012}.

Figure~\ref{HRD} has a few characteristics that are noteworthy. First of all this figure emphasizes again that we need asteroseismology to distinguish between Hydrogen-shell burning (RGB) and Helium-core burning (RC) stars, because both types of stars can occupy the same location in an H-R diagram. Secondly, Fig.~\ref{HRD} shows that for stars above the RC it is much more difficult to determine the period spacings and thus the evolutionary phase. This is in part due the fact that stars high on the RGB oscillate with longer periods and at these lower frequencies the frequency resolution of the data becomes a limiting factor. Additionally the coupling between the p- and g- mode cavity becomes less strong which reduces the number of mixed modes visible at the surface. However this is not true for stars just above the RC. The reason for the non-detections of the period spacings for these stars is at least partly due to rotation \citep{mosser2012mm}.

\subsection{Extraction of global oscillation parameters}
The global seismic parameters $\Delta\nu$ and $\nu_{\rm max}$ are derived from the data using three different methods:
\begin{itemize}
\item CAN: $\Delta\nu$ is obtained from fitting a sequence of Lorentzian profiles spanning three radial orders to the background corrected Fourier power spectrum. This method only considers the central part of the oscillation frequency range and is referred to as `local' method.
$\nu_{\rm max}$ is defined as the centroid of a Gaussian profile fitted on top of two Harvey-like background components in the Fourier power spectrum \citep{kallinger2010,kallinger2012}. In this determination of $\nu_{\rm max}$ the full frequency range is considered in the fitting and hence this is referred to as a `global' approach. 
\item COR: $\Delta \nu$ is obtained from the envelope autocorrelation function (EACF) of the time series \citep{mosser2009} and updated using the universal pattern \citep[UP,][]{mosser2011}. This method takes a relatively large frequency range into account and is referred to as `global' method.$\nu_ {\rm max}$ is obtained as the centre of a Gaussian fit on top of a background computed as the mean slope in log-log \citep{mosser2012}. The background is computed based on relatively narrow frequency intervals bracketing the frequencies at which oscillations have been detected. Therefore, for $\nu_{\rm max}$ this is referred to as a `local' approach. 
\item OCT: $\Delta \nu$ is obtained from the power spectrum of the power spectrum. This method is in between the COR and CAN methods in the sense that it probes a narrower frequency range than COR, but a wider frequency range than CAN. $\nu_{\rm max}$ as the centroid of a Gaussian fit through a smoothed Fourier power spectrum on top of a background first computed with one Harvey-like background component and subsequently improved using the mean slope in log-log \citep{hekker2010}. Because for the initial step the full frequency range was taken into account in the background fitting and in the second step an optimization using relatively narrow frequency intervals bracketing the frequencies at which oscillations have been detected, this $\nu_{\rm max}$ is referred to as a `semi-global' approach. 
\end{itemize}
The differences in the resulting values for $\Delta\nu$ from the local and global approach are significant and allow one to distinguish between red-giant branch stars and red-clump stars \citep{kallinger2012,hekker2012}. The differences in $\nu_{\rm max}$ seem more homogeneously distributed.

\subsection{Surface gravity}
The surface gravity can be computed from the scaling relations (Eq.~\ref{numax} and \ref{dnu}) directly or by using grid-based modeling. The grid-based modeling is performed by two independent implementations based on the recipe described by \citet{basu2010}. One implementation uses BaSTI models \citep{cassisi2006}. The other implementation uses YY isochrones \citep{demarque2004}, models constructed with the Dartmouth Stellar Evolution Code \citep{dotter2007} and the model grid of \citet{marigo2008}. 

\citet{gai2011} already showed that asteroseismic $\log(g)$ is largely model independent. This is confirmed in this study and the results of the different grids are primarily used to validate the results.

\subsection{Solar reference values}
In both the direct method and in the grid-based modelling the solar values of $\nu_{\rm max}$ and $\Delta\nu$ are used. We analysed one year of solar data from the green SPM channel of SOHO/VIRGO \citep{frohlich1997} with the three methods CAN, COR and OCT and find the following: 
\begin{itemize}
\item CAN: $\Delta\nu=134.88 \pm 0.04~\mu$Hz; $\nu_{\rm max}=3120 \pm 5~\mu$Hz \citep{kallinger2010}.
\item COR: $\Delta\nu=134.9 \pm 0.1~\mu$Hz;  $\nu_{\rm max}=3060 \pm 10~\mu$Hz. Note that only the EACF method can be applied to the solar data. 
\item OCT: $\Delta\nu=135.03 \pm 0.07~\mu$Hz and $\nu_{\rm max}=3140 \pm 13~\mu$Hz
\end{itemize}
The solar reference values for $\nu_{\rm max}$ obtained with the different methods are not consistent with each other within 1-sigma. For $\Delta\nu$ the CAN and COR values are consistent with each other within 1-sigma, while this is not the case for the OCT value. Note that the various numbers are formally different but still well within any 3-sigma limit. We expect the main sources for the different solar values for $\Delta\nu$ and $\nu_{\rm max}$ obtained with the different methods to lie in the fact that different definitions in determining the values are used. The computation of a mean value of $\Delta\nu$ is sensitive to the frequency range that is taken into account. The observational definition of $\nu_{\rm max}$ is also different in different methods and depends on whether smoothing is applied or not. Furthermore, $\nu_{\rm max}$ is sensitive to the fitted background.

We investigate the impact of these differences, i.e., we analyse the data for $\log(g)$ using $\Delta\nu$ and $\nu_{\rm max}$ from CAN, COR and OCT with method specific solar reference values obtained from VIRGO data as well as with a so-called intermediate canonical solar reference value: $\Delta\nu=135.1~\mu$Hz and $\nu_{\rm max}=3090~\mu$Hz) \citep{huber2011,huber2013}. The results of these tests are shown and discussed in Sects. 5 and 6.

\section{Tests applied}

For testing the impact on $\log(g)$ of differences in $\Delta\nu$ and $\nu_{\rm max}$ we have computed surface gravities using the global seismic parameters from CAN, COR and OCT using the method specific solar reference value from VIRGO data or the canonical solar value. We also computed values for the surface gravities with and without the correction to the $\Delta\nu$ scaling relation proposed by \citet{white2011}. This results in the following tests:

\begin{itemize}
\item Test 1: grid-based modeling using the original scaling relations and the canonical solar reference values;
\item Test 2: grid-based modeling using the original scaling relations and method specific solar reference values;
\item Test 3: grid-based modeling using the scaling relation for $\Delta\nu$ adapted as suggested by \citet{white2011} and canonical solar reference values;
\item Test 4: grid-based modeling using the scaling relation for $\Delta\nu$ adapted as suggested by \citet{white2011} and the method specific solar reference values.
\end{itemize}

\begin{figure*}
\begin{minipage}{0.49\linewidth}
\centering
\includegraphics[width=\linewidth]{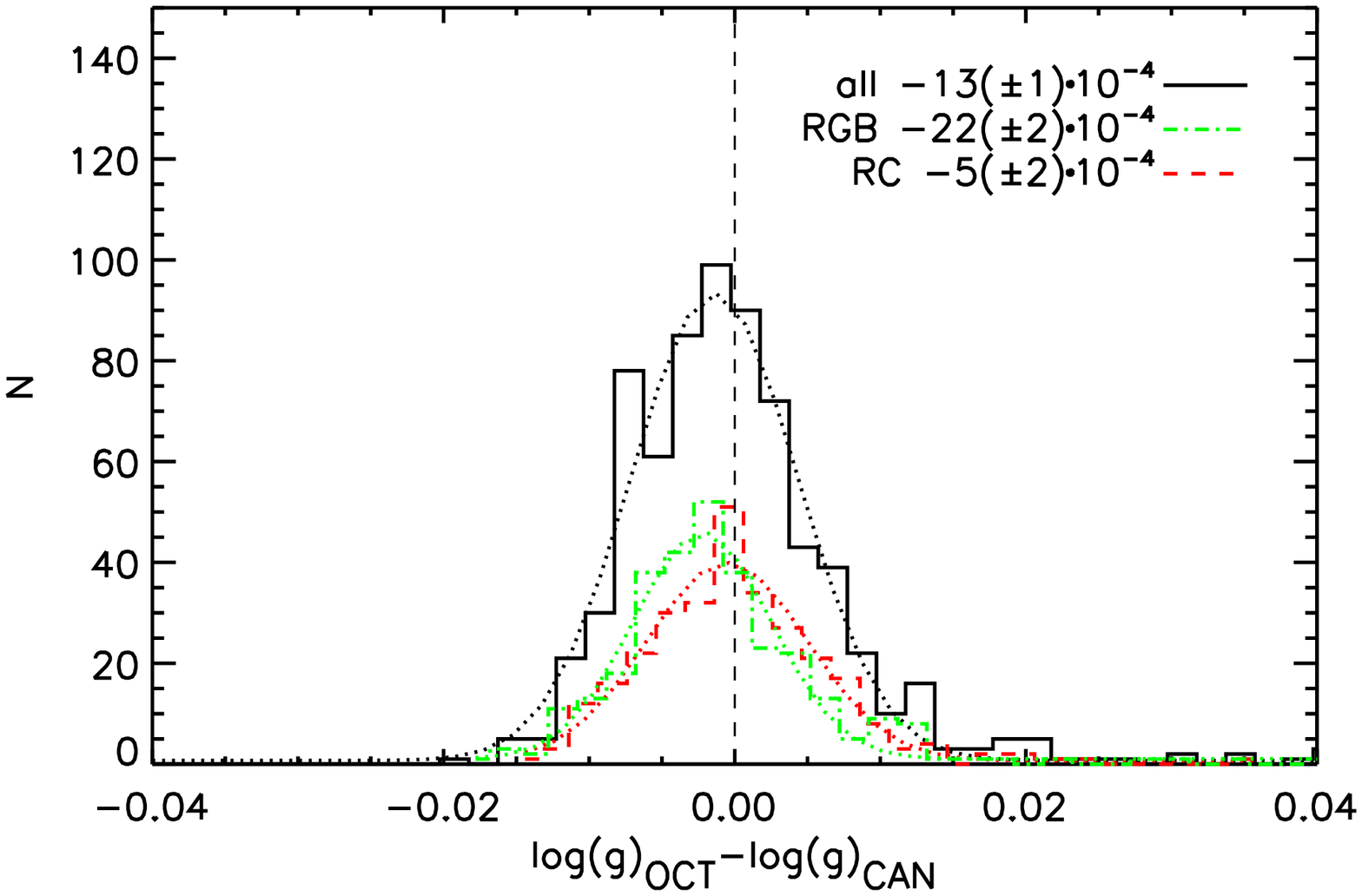}
\end{minipage}
\hfill
\begin{minipage}{0.49\linewidth}
\centering
\includegraphics[width=\linewidth]{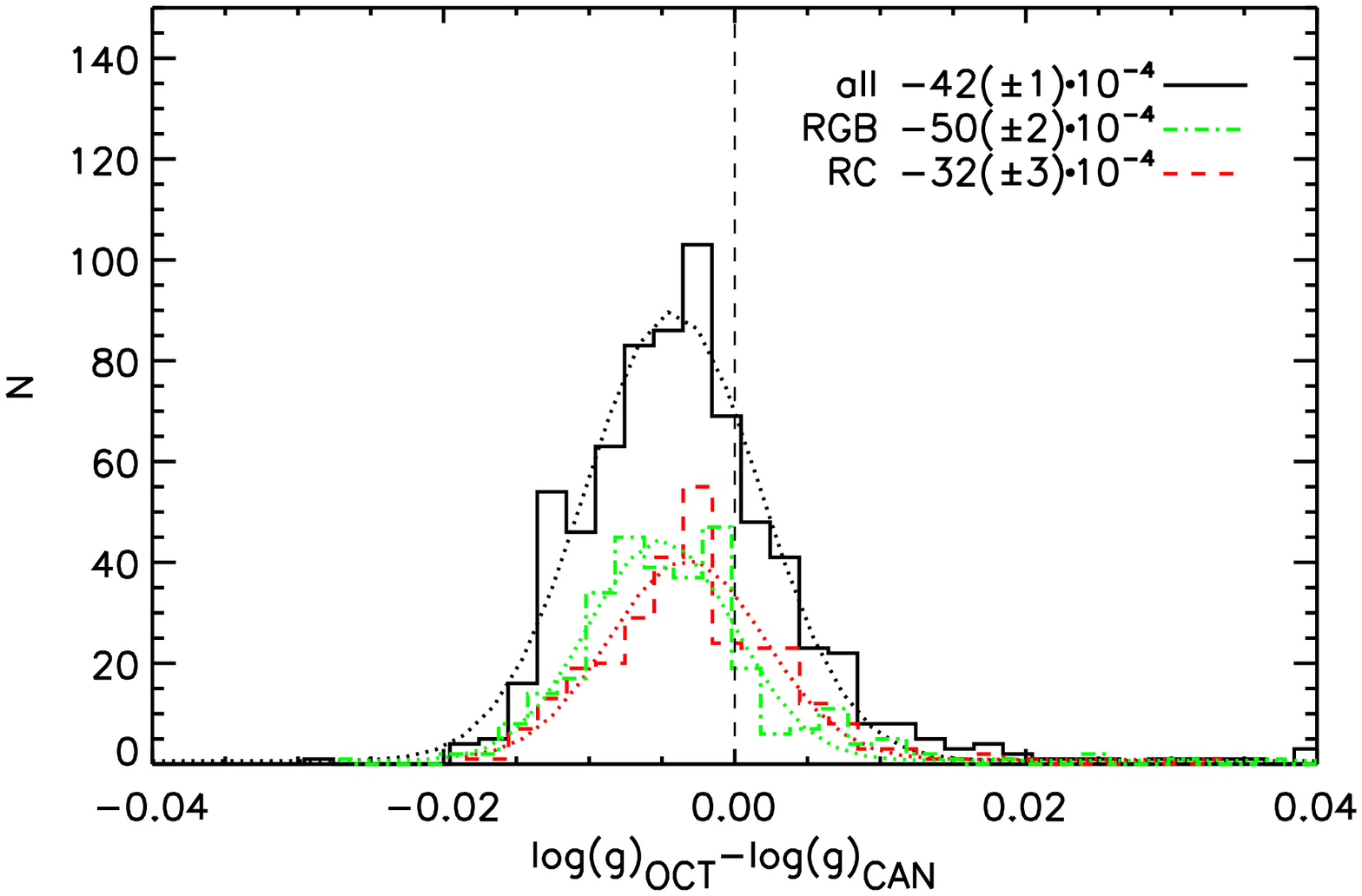}
\end{minipage}
\hfill
\begin{minipage}{0.49\linewidth}
\centering
\includegraphics[width=\linewidth]{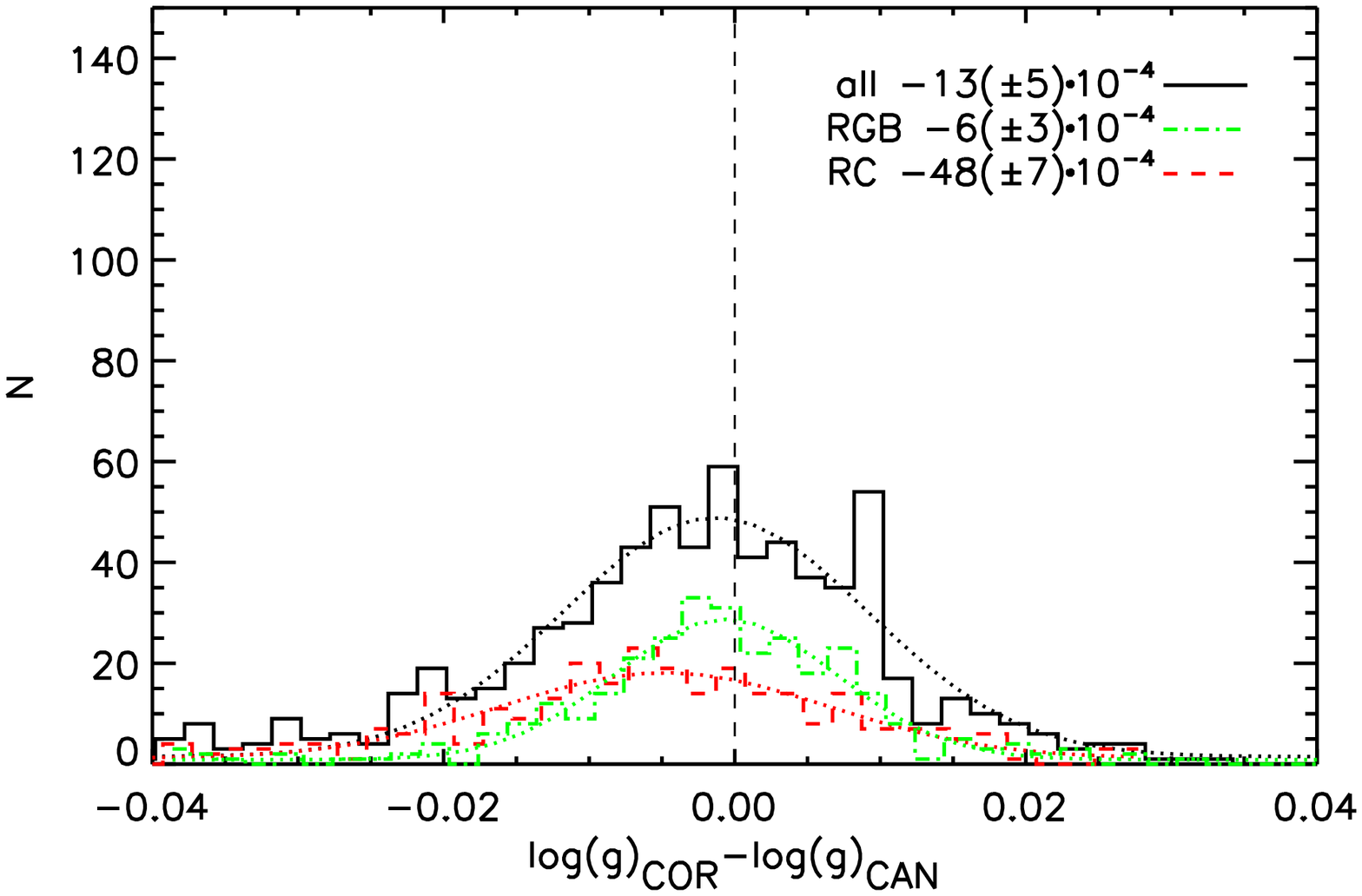}
\end{minipage}
\hfill
\begin{minipage}{0.49\linewidth}
\centering
\includegraphics[width=\linewidth]{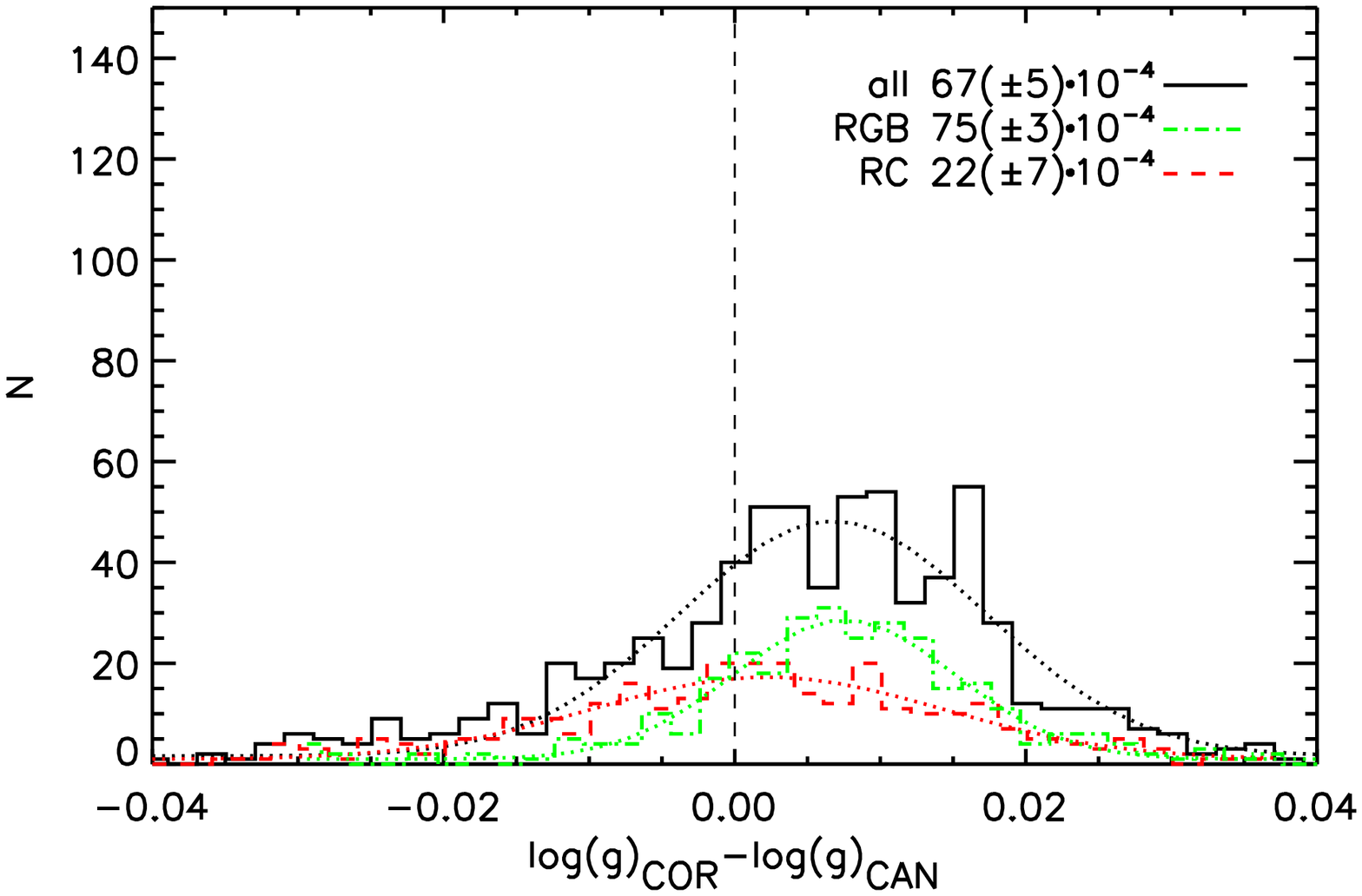}
\end{minipage}
\hfill
\begin{minipage}{0.49\linewidth}
\centering
\includegraphics[width=\linewidth]{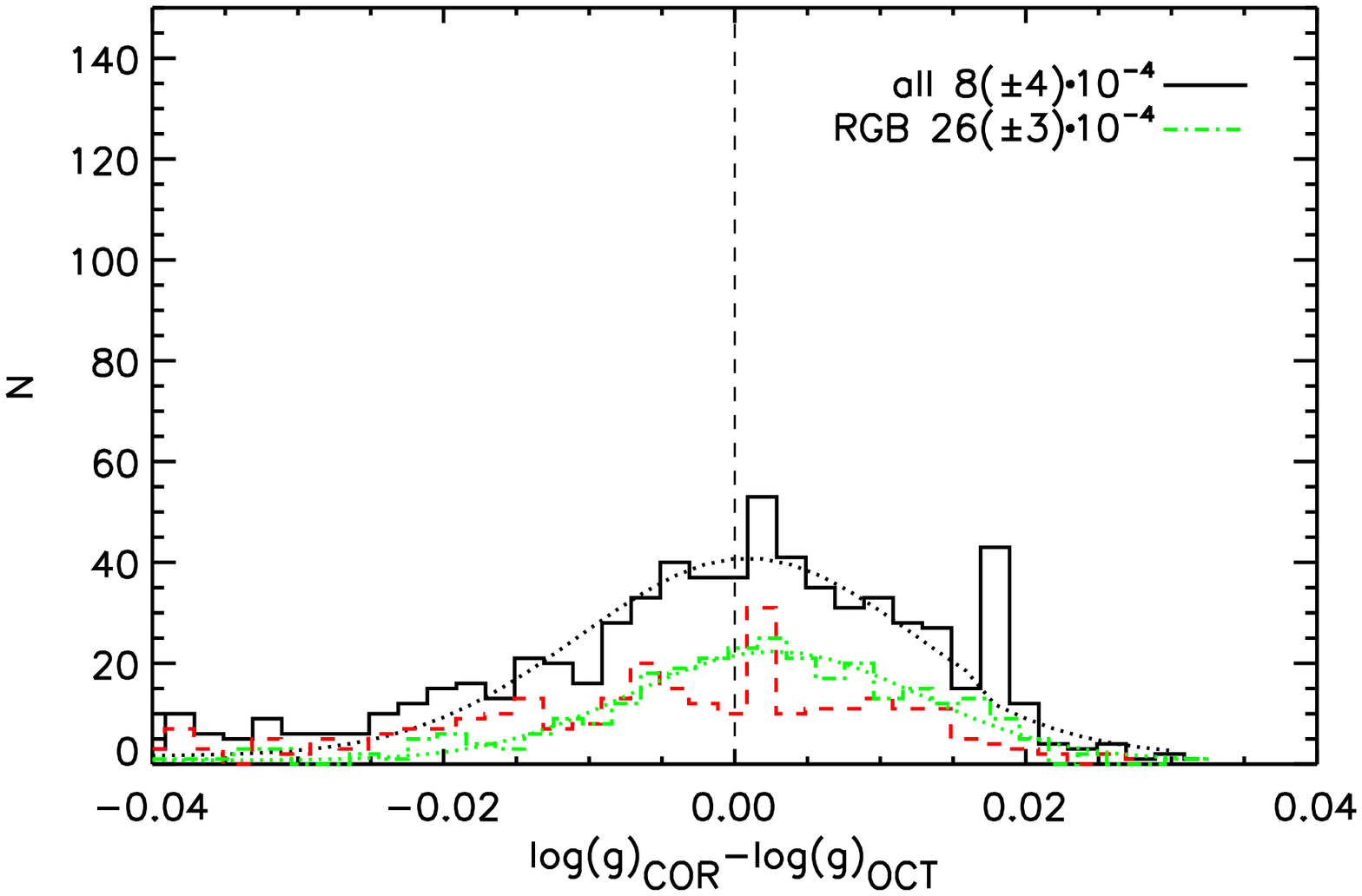}
\end{minipage}
\hfill
\begin{minipage}{0.49\linewidth}
\centering
\includegraphics[width=\linewidth]{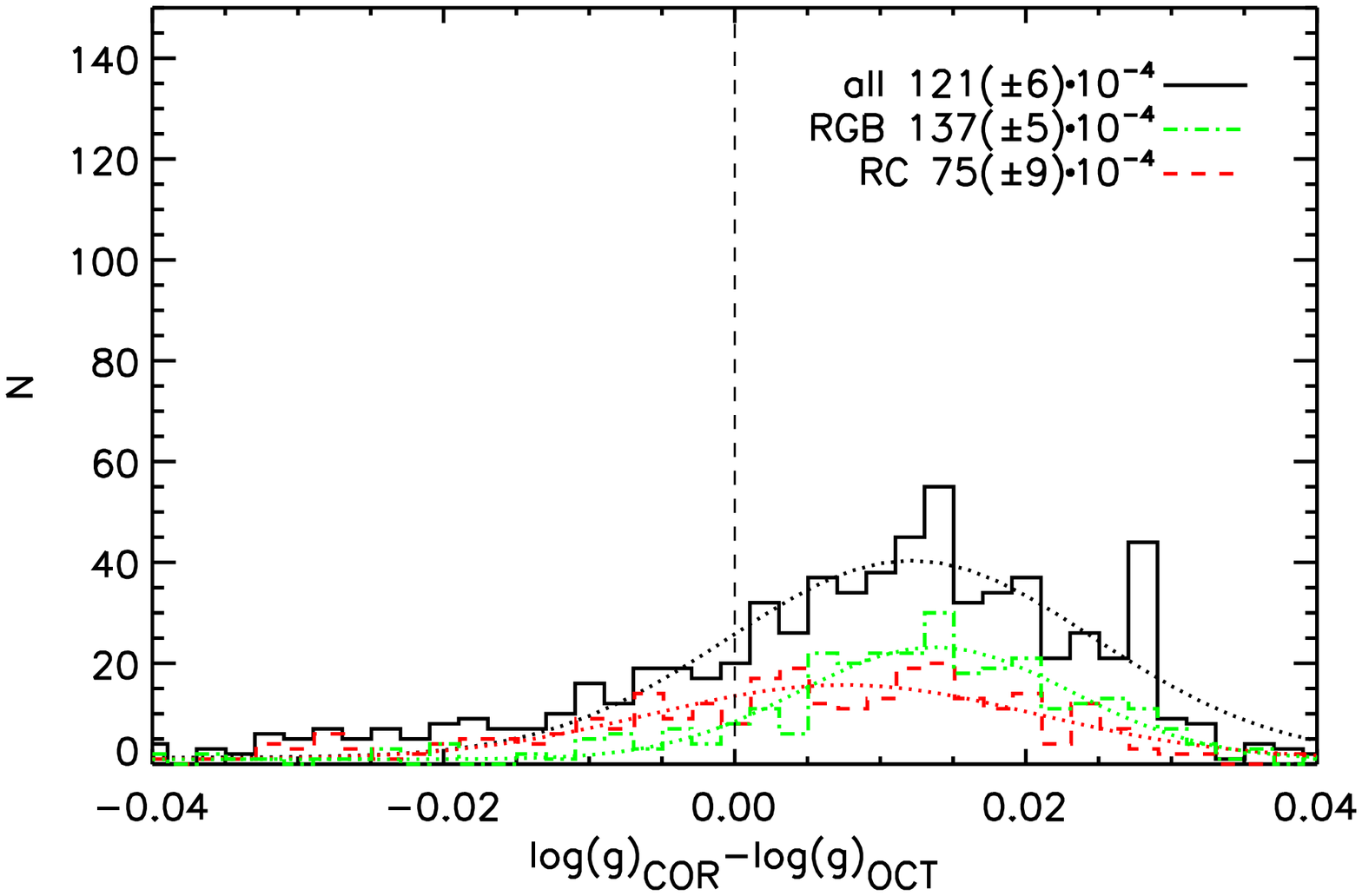}
\end{minipage}
\hfill
\caption{Histograms of the difference in $\log(g)$ obtained with different methods (top row: OCT-CAN, centre row: COR-CAN, bottom row: COR-OCT) and with different solar reference values (left: canonical solar reference value (test 1), right: method specific solar reference value (test 2)). The black solid line indicates the complete sample, the red-dashed line the red-clump stars and the green-dashed-dotted line stars on the red giant branch.  Note that not for all stars we have an evolutionary phase determined. The dotted lines show Gaussian fits to the distributions. The central value and formal 1$\sigma$ uncertainties are given in the legend of each panel. Note that a Gaussian fit through the RC data in the lower left panel did not give a proper representation of the distribution and is hence omitted. The vertical dashed line indicates zero difference.}
\label{logghisto}
\end{figure*}

\begin{figure}
\begin{minipage}{\linewidth}
\centering
\includegraphics[width=\linewidth]{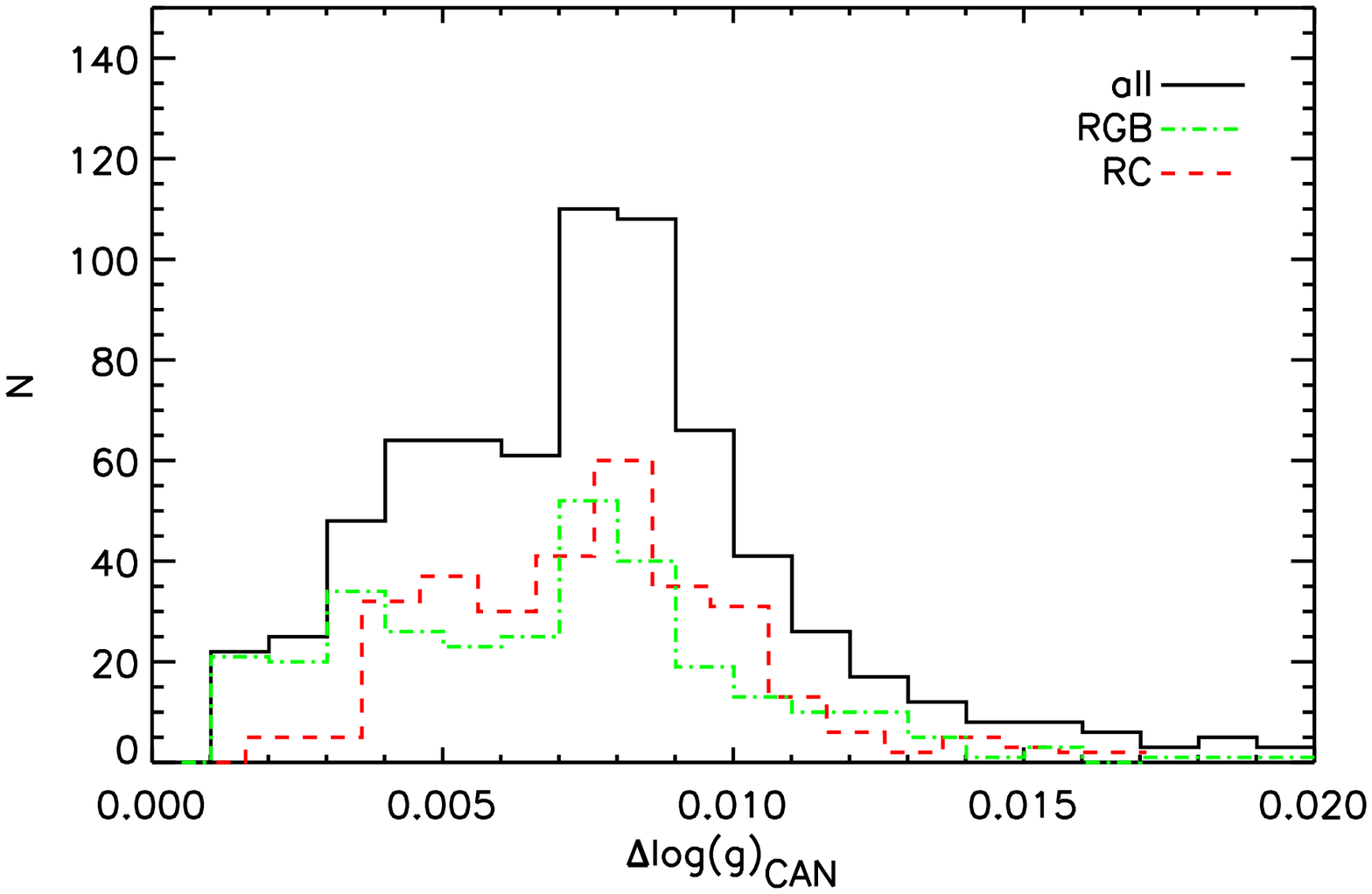}
\end{minipage}
\hfill
\begin{minipage}{\linewidth}
\centering
\includegraphics[width=\linewidth]{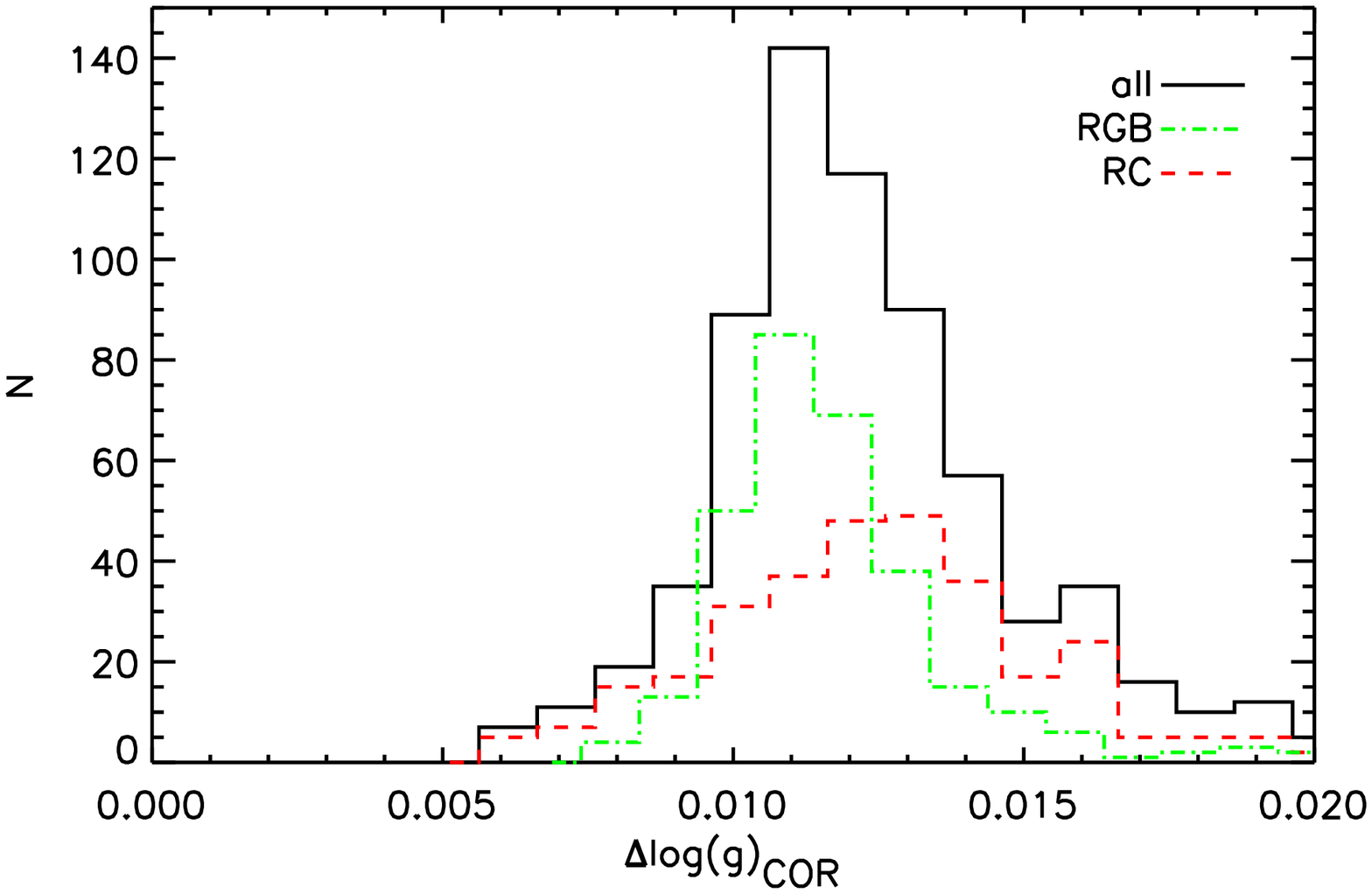}
\end{minipage}
\hfill
\begin{minipage}{\linewidth}
\centering
\includegraphics[width=\linewidth]{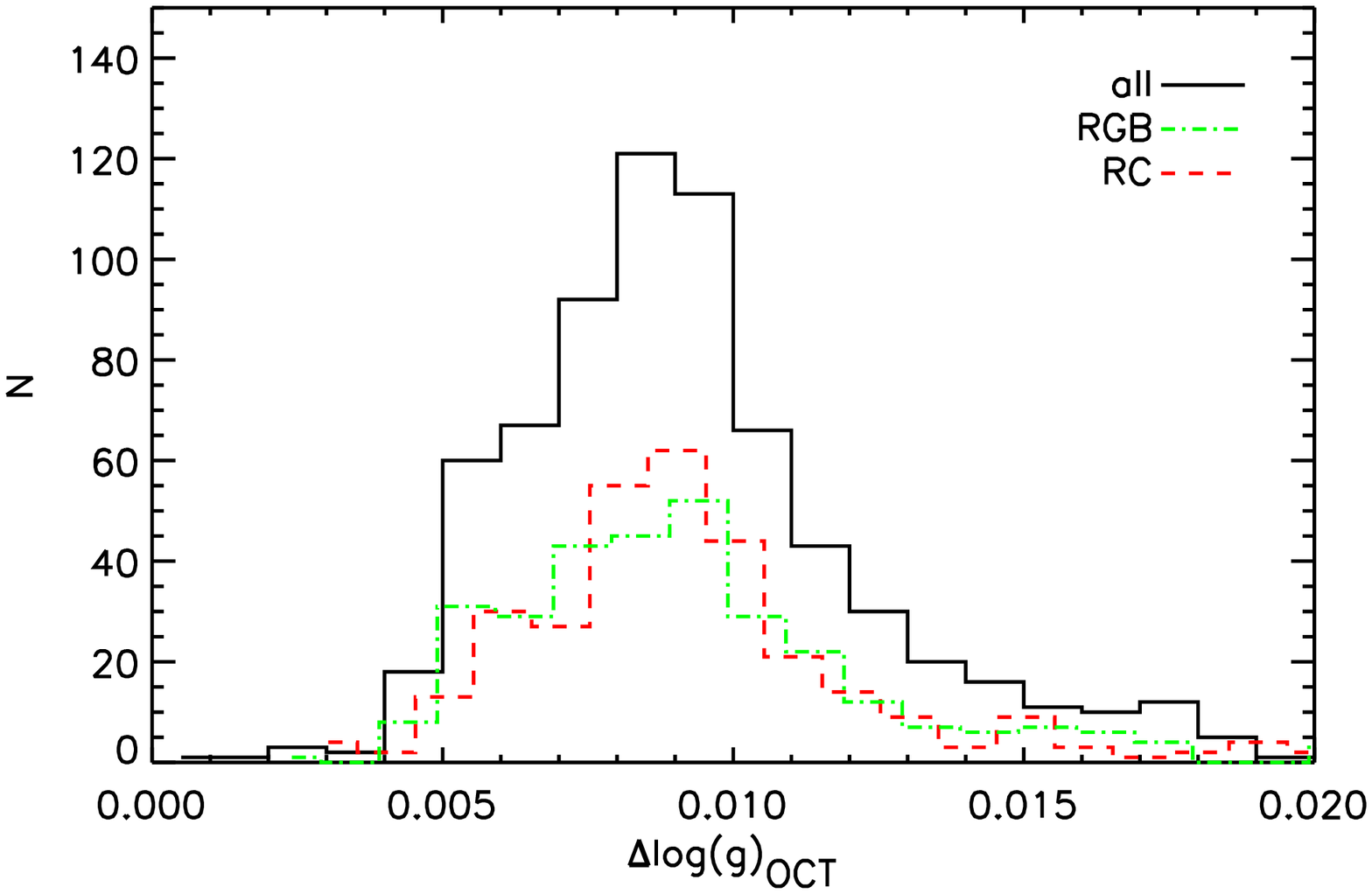}
\end{minipage}
\hfill
\caption{These histograms show the distribution of the uncertainties in $\log(g)$ using global oscillations from CAN, COR and OCT (top to bottom) and canonical solar reference values (test 1). The colour-coding is the same as in Fig.~\ref{logghisto}.}
\label{errlogghisto}
\end{figure}


\begin{figure}
\begin{minipage}{\linewidth}
\centering
\includegraphics[width=\linewidth]{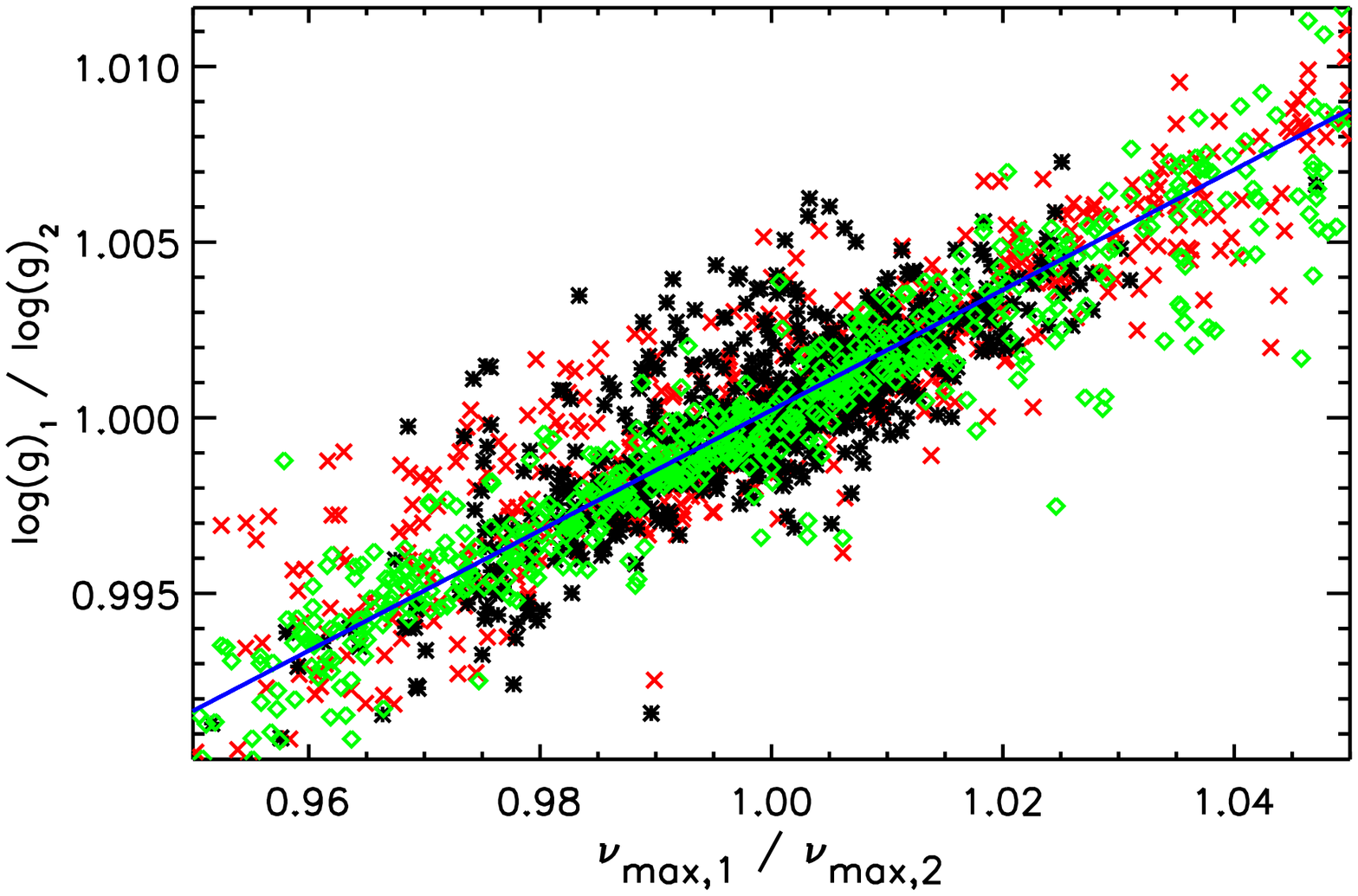}
\end{minipage}
\caption{Ratios of $\nu_{\rm max}$ values from different methods vs. $\log(g)$ values derived from these respective $\nu_{\rm max}$ values. Ratios of different methods are indicated with different colors: black asterisks indicate CAN / OCT, red crosses indicate CAN / COR and green diamonds indicate OCT / COR. The blue solid line is a fit to all results.}
\label{ratio}
\end{figure}

\section{Results}
\subsection{Direct method}
Given Eq.~\ref{numax}, we expect a $\sim0.4$\% change in $\log(g)$ upon a 1\% change in $\nu_{\rm max}$. In the next section we explore the sensitivity of grid-based search methods to changes in both $\Delta\nu$ and $\nu_{\rm max}$ in the range 5~$\mu$Hz $<\nu_{\rm max} <$~250~$\mu$Hz.

\subsection{Grid modeling}
The results of the different experiments as listed in the previous section are shown in Fig.~\ref{logghisto}. These are histograms of the differences in obtained $\log(g)$ values from grid-based modeling using global oscillation parameters derived using different data analysis methods CAN, COR and OCT.

The two columns in Fig.~\ref{logghisto} show the results of test 1 and test 2 respectively (see Sect. 4). Each row shows the difference between two of the methods.  The distributions in the top panels are higher and narrower compared to the distributions in the middle and bottom row. This shows that the CAN and OCT approach for the determination in $\nu_{\rm max}$ give more similar results in $\log(g)$ than the approach adopted by COR. See also Table~\ref{results} for the relative systematics of the results.

Going from the left to the right panels of Fig.~\ref{logghisto}, the canonical solar reference values (test 1) are changed to the method specific solar reference values obtained from the analysis of VIRGO data (test 2). It is clear that the use of different solar reference values introduces a bias in the determinations of $\log(g)$. The shifts are consistent with the difference in solar reference values used for $\nu_{\rm max}$. In other words, it follows from the scaling relation that $\Delta \log(g) \propto \log (\nu_{\rm max \odot,1}/ \nu_{\rm max \odot,2})$. However, it is well known that different methods produce different outputs for the solar values. Therefore, we had expected that the difference in $\log(g)$ would be reduced when using the solar reference values and observed stellar values obtained using a given method (test 2, right column). Evidently, this is not the case. The difference in $\log(g)$ values is significantly smaller when the same solar reference values are used (test 1, left column). This essentially shows that the relative difference in obtained solar values with the different methods is significantly larger than the relative difference in $\nu_{\rm max}$ obtained for red giants between each of the methods.

The left-hand histograms in Fig.~\ref{logghisto} (test 1) also show that there are always offsets between the RGB and RC distributions. We find that for RGB stars the lowest $\log(g)$ is obtained with global oscillation parameters from OCT, and slightly higher values from COR and CAN. For RC stars the distributions are less well defined, making it difficult to be quantitative. 

The uncertainties in $\log(g)$ (Fig.~\ref{errlogghisto} and Table~\ref{results}) show distributions that peak at around 0.01 dex. These distributions are similar for all four tests. In general it seems as if the uncertainty distribution of the RGB stars peaks at slightly higher uncertainties than for RC stars. For CAN and OCT the uncertainties in the RC stars show a wider, flatter distribution compared to the uncertainties of COR indicating that the COR uncertainties are more consistent, albeit slightly higher, than for the other methods. 

\begin{figure}
\begin{minipage}{\linewidth}
\centering
\includegraphics[width=\linewidth]{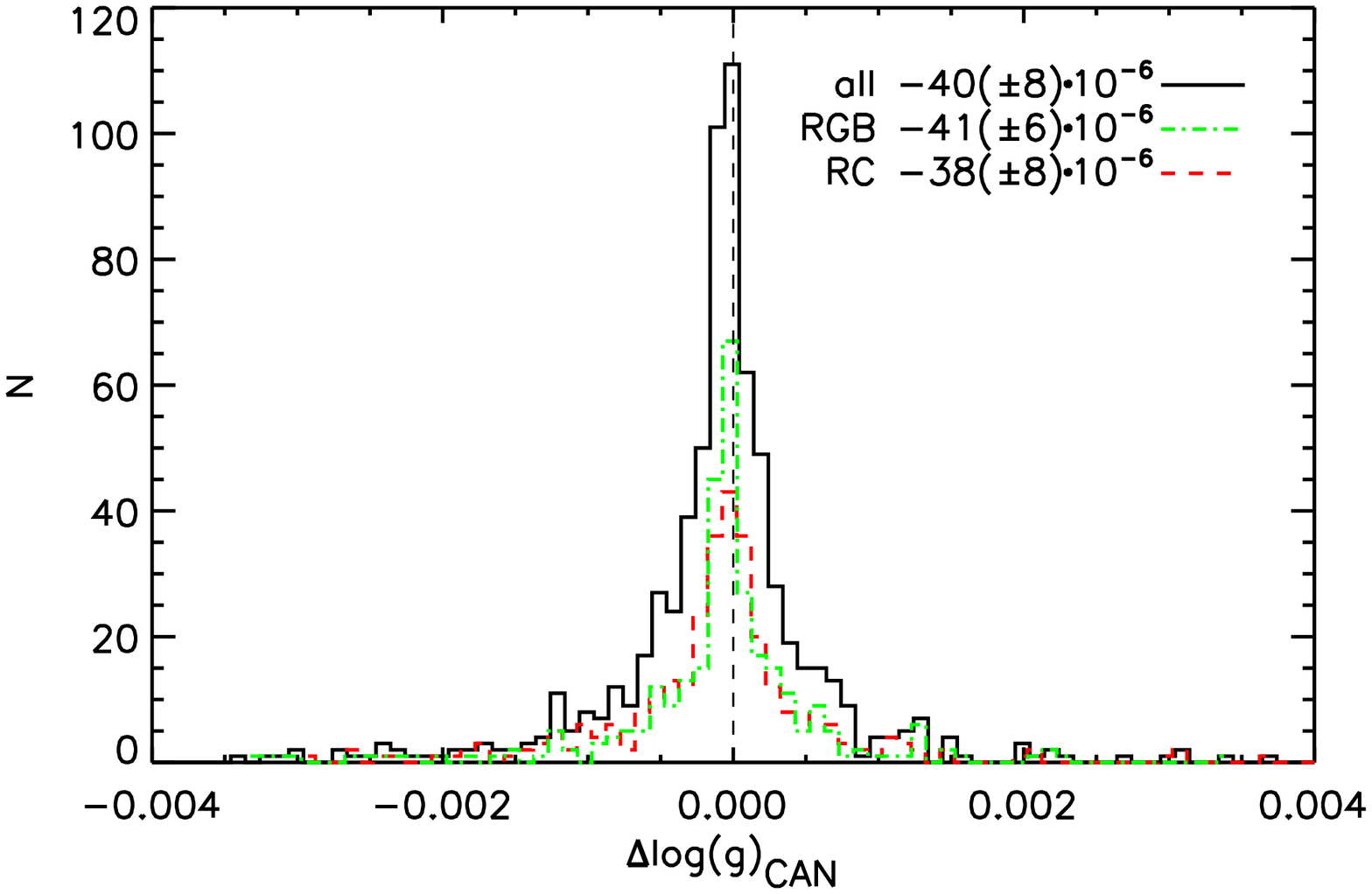}
\end{minipage}
\hfill
\begin{minipage}{\linewidth}
\centering
\includegraphics[width=\linewidth]{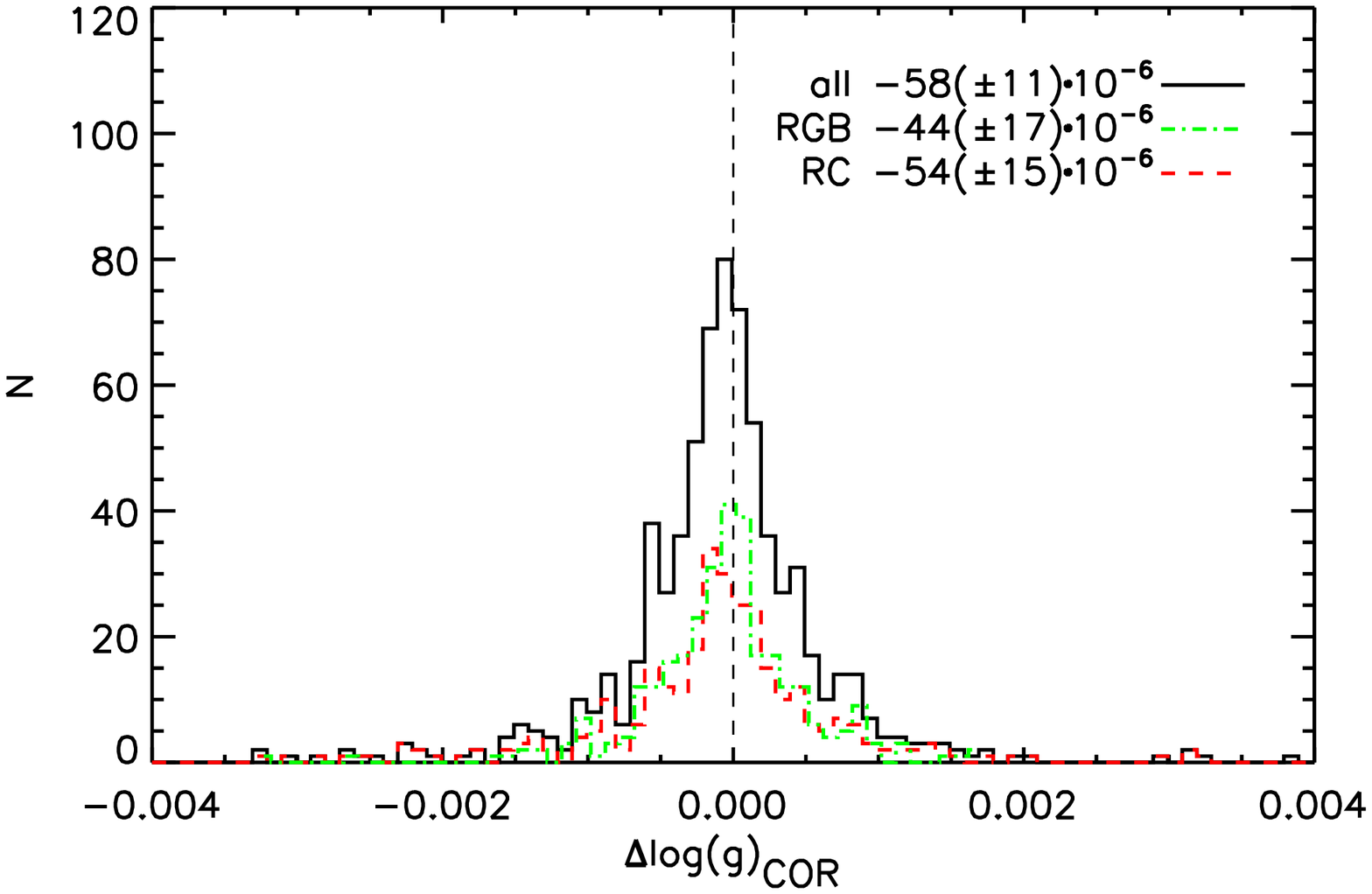}
\end{minipage}
\begin{minipage}{\linewidth}
\centering
\includegraphics[width=\linewidth]{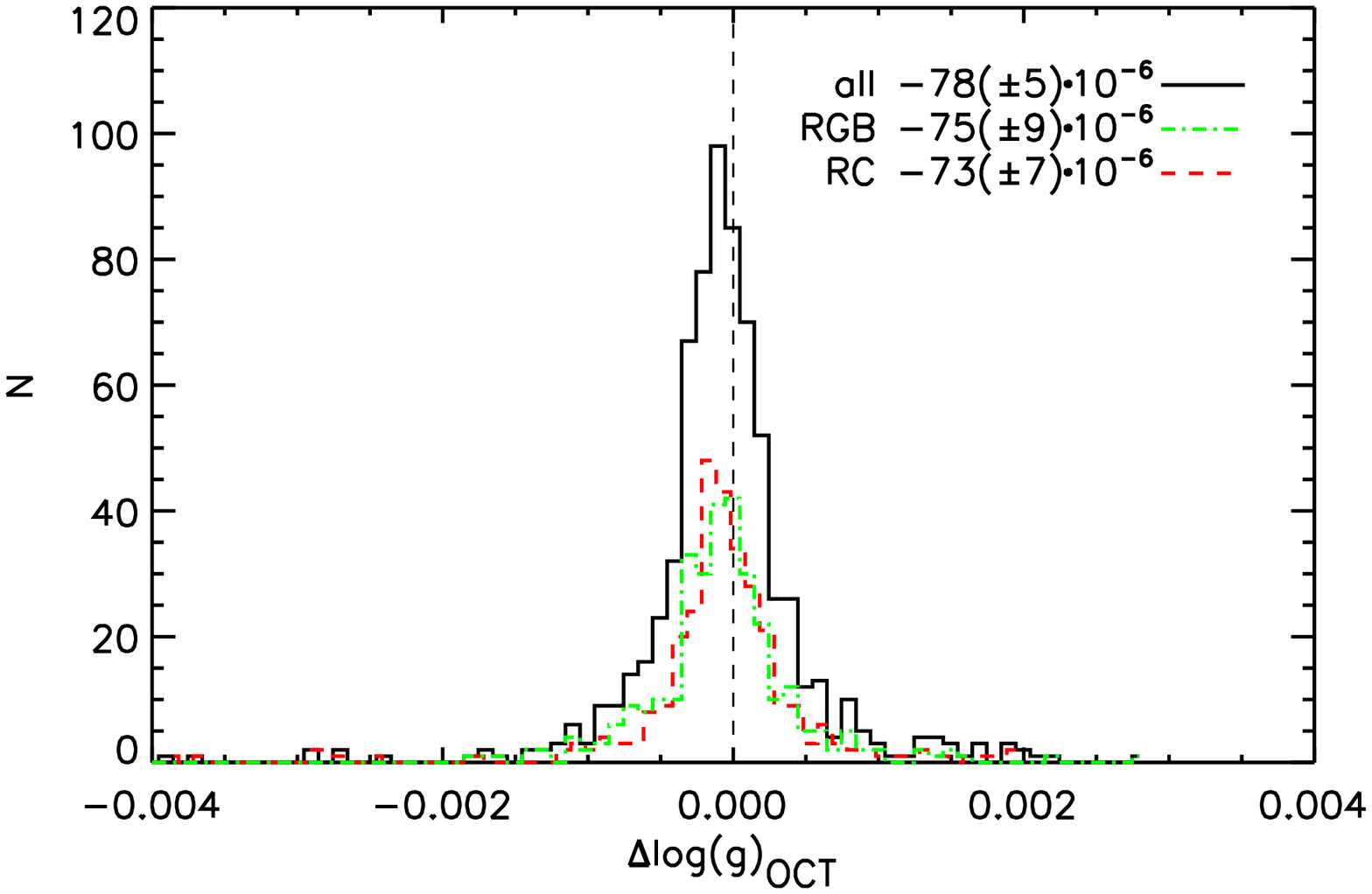}
\end{minipage}
\hfill
\caption{Histograms of the difference in $\log(g)$ obtained using grid-based modeling with the same method, but with and without the correction in the $\Delta\nu$ scaling relation \citep{white2011}. From top to bottom: CAN, COR and OCT. The central values of the distributions and formal 1$\sigma$ uncertainties are given in the legend of each panel. The vertical dashed line indicates zero. The colour-coding is the same as in Fig.~\ref{logghisto}. Note that the horizontal scale has been expanded.}
\label{methhisto}
\end{figure}

As indicated earlier, we also investigate the correction to the $\Delta\nu$ scaling relation by \citet{white2011}. The difference between the results with and without the correction for a specific method are shown in Fig.~\ref{methhisto}. This figure shows that indeed the impact of the 2-3\% correction in $\Delta\nu$ on $\log(g)$ is of the order 0.001, which is well within the uncertainties of the results (see Fig.~\ref{errlogghisto}). This is consistent with what we expect from Eq.~\ref{numax}. We note that a change of a few percent in $\Delta\nu$ will have a significant effect on the determination of the mass and radius.

To investigate the improvement in accuracy of $\log(g)$ obtained from grid-based search methods compared to asteroseismic scaling relations, we show the ratio of $\nu_{\rm max}$ obtained with different methods vs the ratio of $\log(g)$ based on the respective $\nu_{\rm max}$ values (see Fig.~\ref{ratio}). We fit a straight line through the ratios and find a slope of $0.171\pm0.001$. This uncertainty indicates a one-sigma uncertainty estimate of the slope. This indicates a change of 0.171\% in $\log(g)$  upon a change of 1\% in $\nu_{\rm max}$. This is a significantly lower sensitivity than $\sim0.4$\% change in $\log(g)$ upon a 1\% change in $\nu_{\rm max}$ expected from scaling relations. We attribute this to the inclusion of additional constraints in grid-based search methods.

\begin{table*}
\begin{minipage}{\linewidth}
\caption{Summary of ensemble systematics and uncertainties for $\log(g)$ using $\nu_{\rm max}$ and $\Delta\nu$ obtained with different methods using the same canonical solar reference values (test 1).}
\label{results}
\centering
\begin{tabular}{lccccc}
\hline\hline
 & $\log(g)_{\rm OCT} - \log(g)_{\rm CAN}$ & $\log(g)_{\rm COR} - \log(g)_{\rm CAN}$ & $\sigma(\log(g)_{\rm CAN})$ & $\sigma(\log(g)_{\rm COR})$ & $\sigma(\log(g)_{\rm OCT})$\\
 & dex (cgs) & dex (cgs) & dex (cgs) & dex (cgs) & dex (cgs)\\
\hline
All & $-13 (\pm 1) \cdot 10^{-4} $  & $-13 (\pm 5) \cdot 10^{-4} $ & $0.0073 \pm 0.0002$ & $0.0117 \pm 0.0001$ &  $0.0086 \pm 0.0001$ \\
RGB & $-22 (\pm 2) \cdot 10^{-4} $ & $-6 (\pm 3) \cdot 10^{-4} $ & $0.0063 \pm 0.0004$ & $0.0112 \pm 0.0001$ & $0.0086 \pm 0.0001$ \\
RC & $ -5 (\pm 2) \cdot 10^{-4} $ & $-48 (\pm 7) \cdot 10^{-4} $ & $0.0074 \pm 0.0003$ & $0.0124 \pm 0.0001$ & $0.0087 \pm 0.0001$ \\
\hline
\end{tabular}
\end{minipage}
\end{table*}

\section{Discussion}

There is a significant spread in the derived $\log(g)$ values between COR and either CAN or OCT. This large spread indicates not that the COR values are systematically higher or lower than CAN/OCT but that they show a larger scatter with respect to CAN/OCT results. We can understand this as indicating that a less accurate $\log(g)$ is derived when using COR parameters. This could be due to the fact that both CAN and OCT use a more global approach in which they include more prior information to fit the background, while COR uses only a local mean slope in log-log without prior information. Reliable determination of $\nu_{\rm max}$ is dependent on a good determination of the granulation background spectrum. It is possible to argue that the fully global approach of CAN provides the best possible determination of the background and hence of $\nu_{\rm max}$. Ideally one would verify this by computing $\log(g)$ from a Fourier power spectrum obtained from a model for which we know $\log(g)$. Currently the uncertainties in such an approach are too large to have any added value for this analysis. 

There are large biases when using a solar reference value computed with the same method .
We understand these biases as a sign that the current methods are optimised for analysis of frequencies in the red giant regime and not in the solar regime, which leads to relatively large scatter in the solar values compared to the scatter in values determined for red giants. The higher amplitudes of the modes, the lower number of orders as well as the narrow frequency range of the oscillations are likely to improve the consistency between the derived global parameters from different methods.

One might argue that the solar reference is too far off from the frequency regime of the red giants. When it comes to red giants we might therefore consider a `platinum standard', i.e., a reference star that is another red giant. This needs to be a star with extremely well-constrained properties obtained from independent methods. A detached eclipsing binary such as analysed by \citet{hekker2010bin} could be a suggestion. For this star there is an orbital solution, and the mass and radius have been determined accurately (Frandsen, private communication). However, this star has a complicate oscillation pattern and the evolutionary phase has not been determined yet. 

Our suggestion is to derive the most accurate value for $\log(g)$ using oscillation parameters from CAN or OCT with canonical solar reference values. $\log(g)$ from CAN parameters have slightly smaller uncertainties. The drawback of this method that it is relatively time consuming and is not fully automated. $\log(g)$ from OCT parameters has slightly higher uncertainties. Nevertheless, this method shows only small biases in $\log(g)$, which are well within the uncertainties and it is faster than CAN and fully automated. Therefore, we suggest to use oscillation parameters from OCT to obtain a homogeneous analysis. It remains however essential that a selection of methods is applied to validate the results of the chosen method.

We note that there are biases in the distributions between red-clump and red-giant branch stars.  These could be due to the differences arising from a `local' or `global' approach. \citet{kallinger2012,hekker2012} have shown that at least for $\Delta\nu$ it is possible to distinguish between red-clump and red-giant branch stars by the difference in results from the `local' or `global' method. \citet{hekker2012} did not find evidence that $\nu_{\rm max}$ could be used to distinguish between red-clump and red-giant branch stars. However, the results presented here could indicate that it is possible to identify the evolutionary phase of the star from the determination of a `local' or `global' $\nu_{\rm max}$.

In this work we have focussed on the determination of $\log(g)$ for which small changes in $\Delta\nu$ are insignificant. We recognize however that a change of a few percent in $\Delta\nu$ will have a significant effect on the determination of the mass and radius of the star. Using the scaling relations (Eqs.~\ref{numax} and \ref{dnu}) it is straightforward to derive that a 5\% uncertainty in $\nu_{\rm max}$ and 2\% uncertainty in $\Delta\nu$ lead to a $\sim10$\% uncertainty in stellar mass and a $\sim6$\% uncertainty in stellar radius. This is significantly higher than the $\sim2$\% uncertainty in $\log(g)$ upon a 5\% uncertainty in $\nu_{\rm max}$.

\section{Conclusions}
For grid-based modeling we compared the $\log(g)$ values obtained from the seismic parameters derived from the CAN, COR and OCT methods using a grid of BaSTI models. We can draw the following conclusions:
\begin{itemize}
\item $\log(g)$ from oscillation parameters from the CAN and OCT method are more similar and more precise than the results from COR;
\item the use of same canonical solar reference value reduces the biases in $\log(g)$ compared to using a method specific solar reference value (this is at least true for red giants analysed here);
\item there are small biases between the results for red-clump and red-giant branch stars;
\item the uncertainties in $\log(g)$ are of the order of 0.01dex (cgs);
\item the biases due to different methods are within the uncertainties when using the same canonical solar reference value;
\item the correction in the $\Delta\nu$ scaling equation of 2-3\% as proposed by \citet{white2011} does not influence the determination of $\log(g)$ significantly;
\item grid-based search methods show $\log(g)$ to have a lower sensitivity to small changes in $\nu_{\rm max}$ than is apparent from the direct scaling relations.
\end{itemize}


\begin{acknowledgements}
We thank R. A. Garc\'ia (CEA) for all the time and efforts he has put into the correction of the raw \textit{Kepler} data to make them suitable for the analyses presented in this paper. SH acknowledges financial support from the Netherlands Organisation for Scientific Research (NWO). YE and WJC acknowledge support from STFC (The Science and Technology Facilities Council, UK). SB acknowledges NSF grant AST-1105930 and NASA grant 495
NNX13AE70G. TK acknowledges financial support from the Austrian Science Fund (FWF P23608). This work has made use of BaSTI web tools. SOHO is a mission of international collaboration between ESA and NASA
\end{acknowledgements}

\bibliographystyle{aa}
\bibliography{logg}
\end{document}